\documentclass{article}

\usepackage{arxiv}

\usepackage[utf8]{inputenc} 
\usepackage[T1]{fontenc}    
\usepackage{hyperref}       
\usepackage{url}            
\usepackage{booktabs}       
\usepackage{amsfonts}       
\usepackage{nicefrac}       
\usepackage{microtype}      
\usepackage{lipsum}		
\usepackage{graphicx}
\usepackage{natbib}
\usepackage{doi}
\usepackage{placeins}

\begin{document}
\title{Topology of molecular clouds in the PHANGS-JWST catalogue in relation to the physical characteristics of their galaxies.}

\date{May 31, 2025}	
\author{ \href{https://orcid.org/0009-0004-3459-4772}{\includegraphics[scale=0.06]{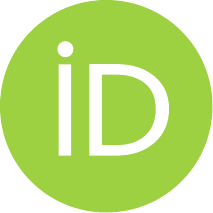}\hspace{1mm}Ignacio ~Morales-Gil}
\\
	Madrid, Spain  \\
	\texttt{ignaciomorgil@gmail.com} \\
}



\renewcommand{\shorttitle}{\textit{arXiv} Template}

\hypersetup{
pdftitle={A template for the arxiv style},
pdfsubject={},
pdfauthor={},
pdfkeywords={molecular clouds, topology, SFR},
}

\maketitle

\begin{abstract}
Molecular clouds are affected by dynamical processes such as cloud merging and stellar feedback that alter their topology. For this reason, the molecular clouds in the PHANGS-JWST catalogue are studied under the light of topological data analysis. This study shows correlations between physical characteristics of galaxies and topological summaries of molecular clouds.

\end{abstract}


\section{Introduction}
The PHANGS (Physics at High Angular resolution in Nearby GalaxieS) program uses the JWST, together with the HST, to study star formation along all of its life cycle in relation to molecular clouds among other topics. Molecular clouds are linked to star formation in galaxies, through dynamical processes that modify the topology of the clouds. Collisions among molecular clouds join and merge components, increasing SFR; and stellar feedback can open holes and voids in the clouds, so the interaction between star formation and molecular clouds has topological consequences for the galaxy. Therefore, a topological analysis can cast light on this interaction.
\FloatBarrier
\begin{figure}[h!]
    \centering
    \includegraphics[width=0.33\linewidth]{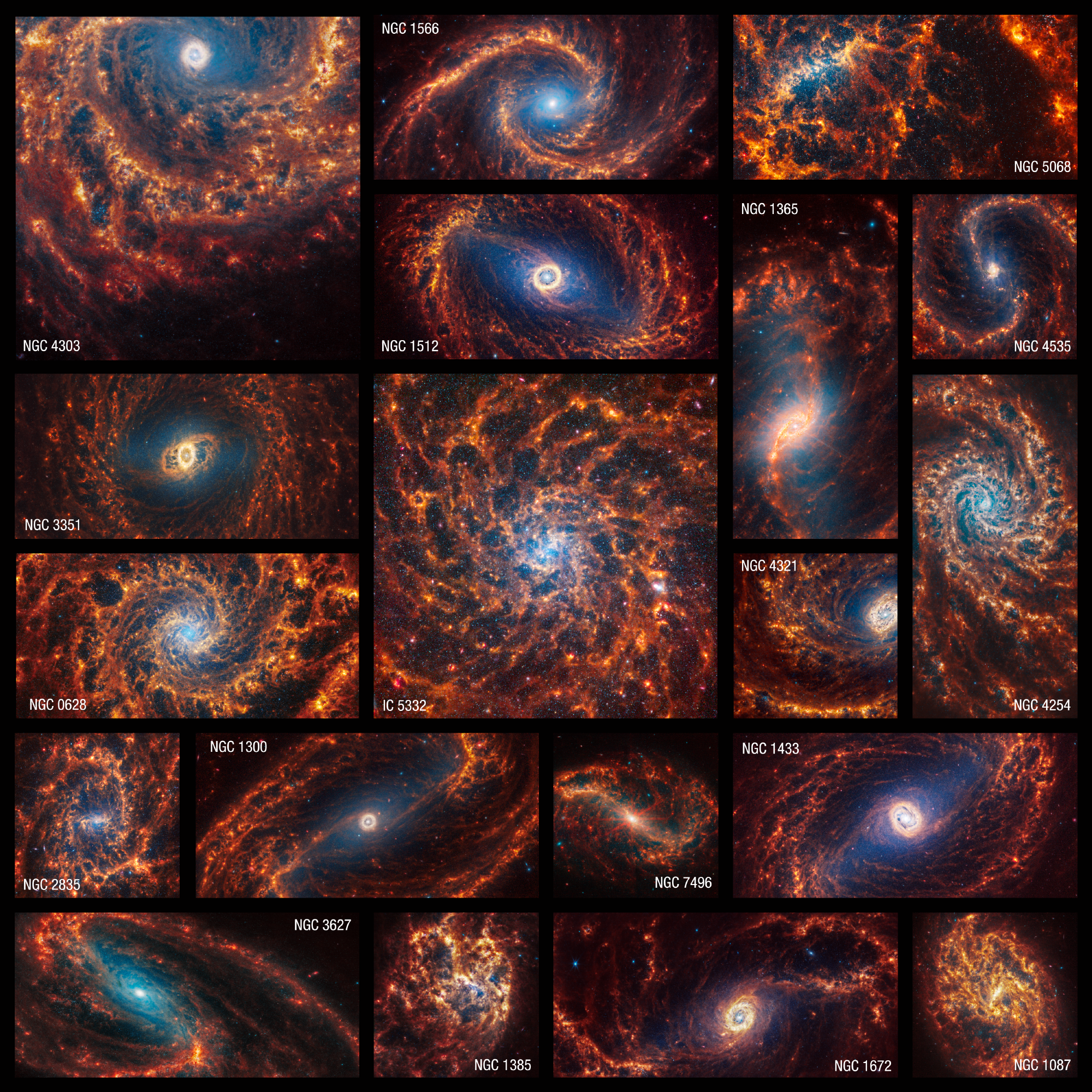}
    \caption{Credit:  NASA, ESA, CSA, STScI, Janice Lee (STScI), Thomas Williams (Oxford), PHANGS Team
Acknowledgement:  Elizabeth Wheatley (STScI)}
    \label{NGC628 example}
\end{figure}
\FloatBarrier
Topological data analysis is a recent branch of math that studies the evolution of complexes in data when varying parameters, hence the name of its main tool: persistence homology. There are several versions of this tool. This work uses cubical persistence homology, more computationally efficient for images, as this work studies maps of molecular clouds as observed by JWST.

In detail, the merging of clouds affects the 0 dimensional homology, concerning connected components; whereas the holes opened by stellar winds are registered by the 1 dimensional homology on a head-on image of the galaxy, that detects rings up to homomorphism.

\section{Methods}
I used images from the PHANGS-JWST catalogue, particularly v1p0. They're available publically at CADC (CANFAR), the source from which I downloaded the data.

\begin{figure}[h!]
    \centering
    \includegraphics[width=0.75\linewidth]{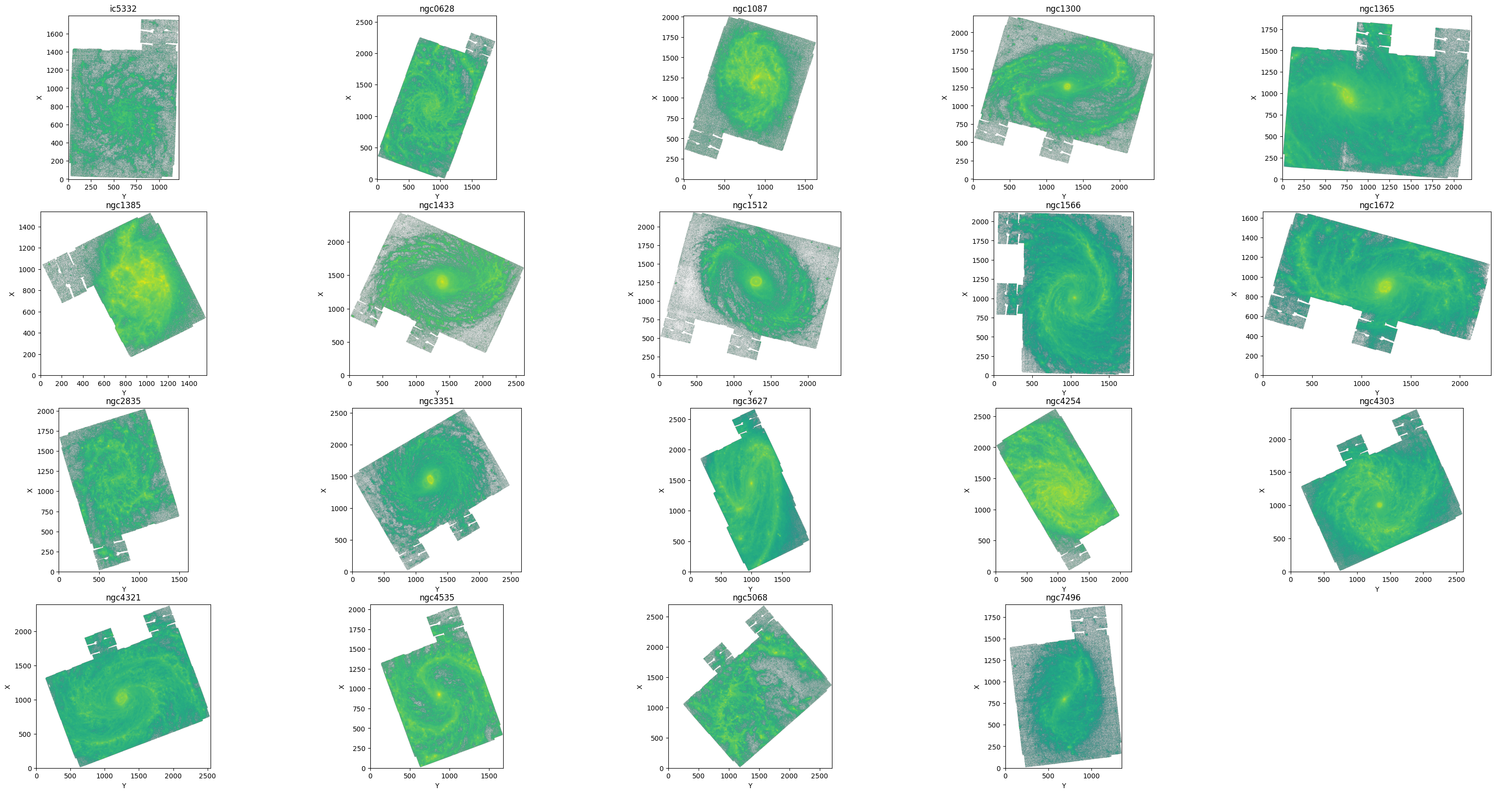}
    \caption{Maps used in this work with logarithmic norm for the color.}
    \label{galaxies}
\end{figure}

\begin{figure}[h!]
    \centering
    \includegraphics[width=0.15\linewidth]{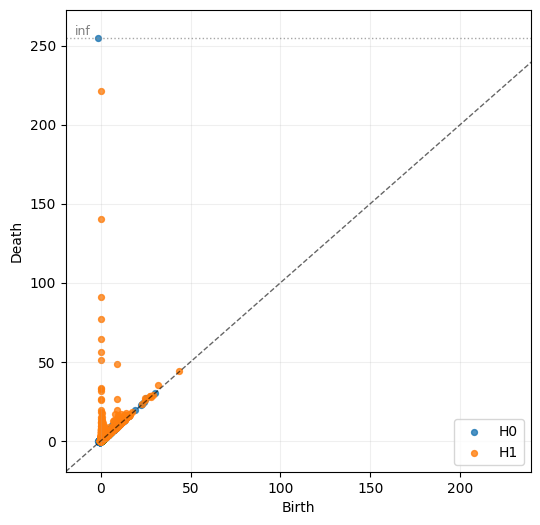}
    \caption{As an example, this figure shows the persistence diagram of NGC 1433 under JWST's f1000w filter.}
    \label{NGC1433_pd}
\end{figure}

Then, for every map, I proceeded with the topological data analysis workflow, consisting of computing the persistence diagrams \ref{NGC1433_pd} of dimensions 0 and 1 with the package cripser \cite{2005.12692}, transforming the output into ripser format \cite{ctralie2018ripser} and computing the topological summaries, namely:
\begin{enumerate}
	\item The sum of longevities of 1 dimensional homology elements, $\sum \rm{LH1}$
	\item  A linear combination of 0 and 1 dimensional statistics: CPH$=2\cdot \rm{var(LH0)}+\frac{\sum \rm{LH1}}{2-\sqrt(2)} $
	\item The sum of longevities of 0 dimensional homology elements,$\sum \rm{LH0}$
	\item The quotient of the sum of longevities of 1 dimensional homology elements and the sum of longevities of 0 dimensional homology elements, $\frac{\sum \rm{LH1}}{\sum \rm{LH0}}$
	\item The quotient of the number of 1 dimensional homology elements and 0 dimensional homology elements, $\frac{H1}{H0}$
	\item The logarithm of the variance of longevities of 0 dimensional homology elements, $\log_{10}(\rm{var(LH0)})$
	\item The logarithm of the variance of longevities of 1 dimensional homology elements, $\log_{10}(\rm{var(LH1)})$
\end{enumerate}

\subsection{Topology under the f1000w filter}

Firstly, all 19 galaxies' maps with the f1000w filter are processed by the TDA workflow described above.
The aim is to study the relationship with physical magnitudes obtained from \cite{lee_phangsjwst_2023}. These are: 
\begin{itemize}
	\item Galaxy stellar mass. Following Leroy et al. (2021), based on Spitzer IRAC 3.6 $\mu$m when available, or WISE 3.4 $\mu$m, and the mass-to-light ratio prescription of Leroy et al. (2019) calculated as a function of radius in the galaxy.
	\item SFR$_{tot}$, the total galaxy SFR. Based on GALEX far-UV and WISE W4 imaging with the SFR prescription calibrated to match results from population synthesis modeling of Salim et al. (2016, 2018) as in Leroy et al. (2021).
	\item Stellar mass effective radius from Leroy et al. (2021) and closely resembling the near-IR effective radius (Munoz-Mateos et al. 2015).
	\item  Integrated CO (2-1) luminosity. Scale by $\alpha_{\rm{CO}}$ $\approx$ 6.7 M$_\odot$ pc$^{-2}$ (K km s$^{-1}$)$^{-1}$ to estimate M$_{\rm{mol}}$ including helium and metals using a fixed CO-to-H2 conversion factor. 
	\item Atomic gas mass, not including helium, from HyperLEDA (Makarov et al. 2014).
	\item Gas phase metallicity on the S-cal system (Pilyugin \& Grebel 2016) estimated at Re by Groves et al. (2023). 
	
\end{itemize}

The plots of physical magnitudes vs topological summaries are shown in the appendix \ref{topology_vs_physics_img_table}.

Afterwards, I analysed the correlations with a Stan \cite{rstan} MCMC model. In particular the model uses 4 chains, 10000 iterations, with 2500 for warmup. The model is robust given that it uses a t-Student distribution and leaves the $\nu$ parameter free.

\subsubsection{Linear fits}
Selecting the variables with significant linear correlation with SFR$_{\rm{tot}}$, I fitted them to a linear model in Stan.

The linear model uses the following priors:
\begin{itemize}
    \item A t-Student prior for the dependent variable, the topological summary
    \item Normal distributions for the linear parameters, with mean and dispersions given by a least squares initial fit
    \item A Cauchy distribution for the dispersion, with $\mu=0$ and $\sigma$ estimated as the error on the inhomogeneous term of the least squares fit
    \item A gamma(2, 0.1) distribution for the degrees of freedom parameter
\end{itemize}

\subsection{Topologies of IC5332}
Then  I analysed IC5332 images with all filters available.
\FloatBarrier
\begin{figure}[h!]
    \centering
    \includegraphics[width=0.75\linewidth]{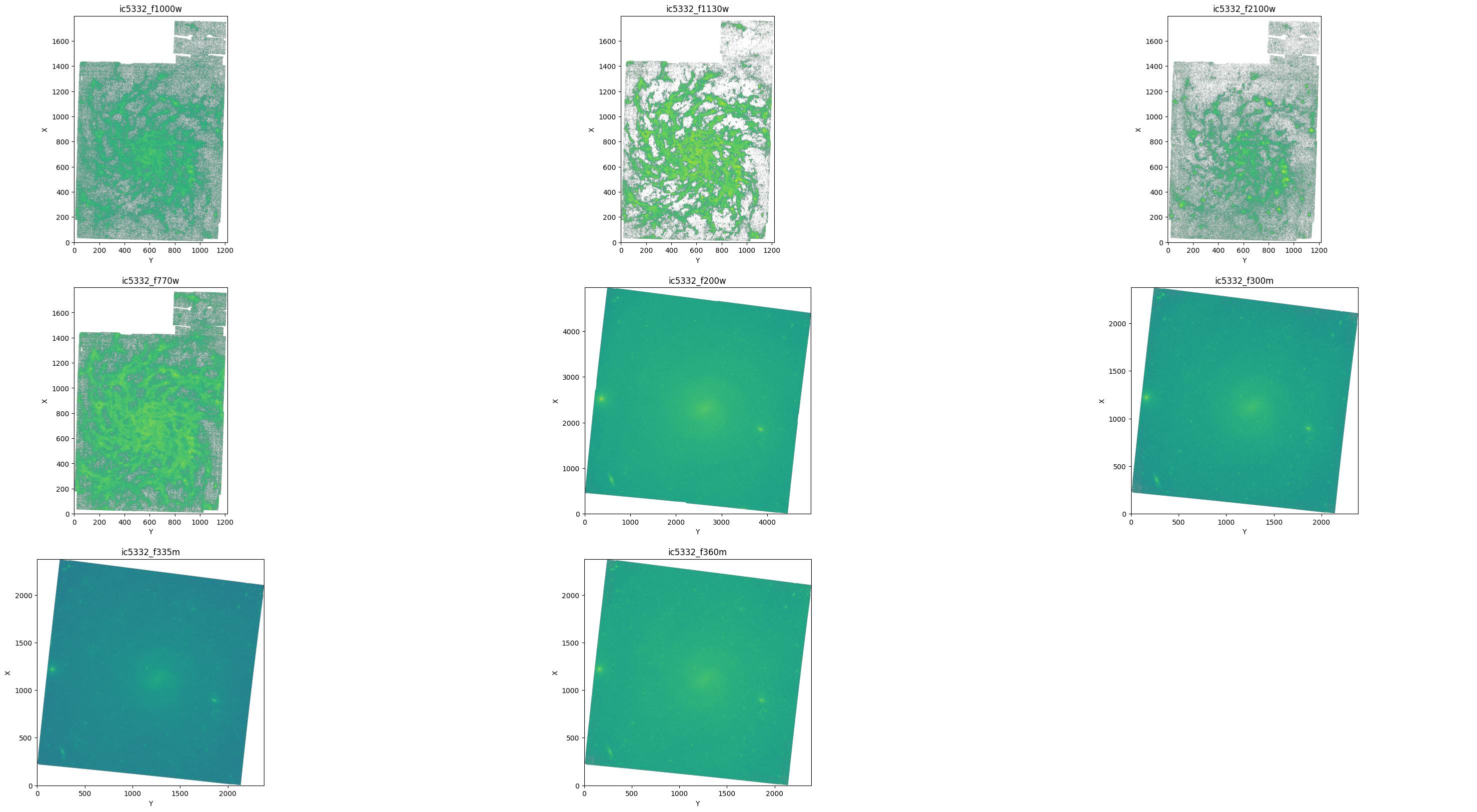}
    \caption{Maps of IC5332 under different filters.}
    \label{IC5332s}
\end{figure}
\FloatBarrier


\section{Results}

SFR$_{tot}$ is the physical magnitude that correlates significantly with the most topological summaries.
Physical to topology correlation table.

The table of posteriors and predictive posterior distributions are in the appendix \ref{regression_posteriors} \ref{DPPs}\\
Linear fits, posteriors and plots are also included in the appendix  \ref{linear_fits}\\

Except for at most a few outliers, the topological summaries are similar across all filters. A table of variability statistics of topological summaries in IC5332 is included next.

\begin{table}[h!]
    \centering
    \begin{tabular}{|c|c|c|c|c|c|c|c|c|c|c|}
    \hline
         &var$(LH_0)$  & $\sum \rm{LH_1}$ & CPH & var$(LH_1)$ & $\sum \rm{LH_0}$& mean$(LH_0)$& mean$(LH_1)$& H$_0$ & H$_1$&$\frac{H_1}{H_0}$   \\
    \hline
    $\sigma$ & 0.022 &  81000& 140000 & 1.4 &24000  & 0.078 &  0.066&0.047 & 570000& 280000\\
    \hline
    $\mu / \sigma$& 0.61 &0.74  & 0.74 & 1.5 & 1.5 & 1.2 & 1.5 &12  &  1.02& 1.1\\
    \hline
    median$/ \sigma$& 0.19 & 0.42 & 0.42 & 1.4 & 1.1 & 0.80 & 2.0 & 12& 0.68 & 0.78\\
    \hline
    \end{tabular}
    \caption{Statistics of topological summaries across filters for IC5332}
    \label{IC5332_estadisticatopologia}
\end{table}

\section{Discussion}
Galaxies were sufficiently head-on to have a view of the holes. The analysis performed in this work cannot be trivially translated to edge-on galaxies.\\
The package cripser detects some creator and destroyer cells outside the map of the galaxy, that is outside of the coverage  of the observations. However, those are minimal in proportion to the total number of cells. 

\begin{table}[h!]
    \centering
    \begin{tabular}{|c|c|c|}
            \hline

        Cell type & Mean proportion & Standard deviation \\
        \hline
        Creator cells for dimension 0 & 1.9 e -5 & 7.9 e -6\\
                \hline

        Destroyer cells for dimension 0 & 3.0 e -3 & 7.7 e-4\\
                \hline

        Creator cells for dimension 1 &  3.0 e -3 & 1.1 e-3\\
                \hline

        Destroyer cells for dimension  1 &  9.6 e-5 & 3.7 e-4\\
                \hline

    \end{tabular}
    \caption{Proportion of cells outside the coverage. The mean and standard deviation are computed over all 19 galaxies in the sample.}
    \label{celulas_fuera}
\end{table}

Regression gives another way to estimate physical parameters. As the explanation of correlations is more easily acceptable with SFR$_{tot}$, I selected it for the linear plots.\\
CPH is dominated by $\sum LH_1$ so I didn't compute the regression of both.\\
Whether a generalization with other filters is possible, is the reason to compute topologies for the same galaxy on different filters. The quotient of mean or median to the standard deviation for the topological summaries in IC5332 is moderate in many cases, which suggests some similarity of the topologies among filters. The same analysis should be performed with others galaxies.\\
The  Wasserstein and bottleneck distances among persistence diagrams are too computationally heavy on memory, so the comparison of topologies is made in terms of topological summaries.

\section{Conclusion}
Correlations among physical and topological characteristics of molecular clouds exist in galaxies. Particularly, some topological summaries of the molecular clouds are useful to predict SFR in a galaxy.\\
Many aspects of galactic physics can be studied with topological tools.
\section{Acknowledgements}
The author acknowledges the use of the Canadian Advanced Network for Astronomy Research (CANFAR) Science Platform operated by the Canadian Astronomy Data Centre (CADC) and the Digital Research Alliance of Canada, with support from the National Research Council of Canada (NRC), the Canadian Space Agency (CSA), CANARIE, and the Canada Foundation for Innovation (CFI).
This article uses the LaTeX template available at https://github.com/kourgeorge/arxiv-style.
I also thank useful comments from Patricia Sánchez Blázquez (UCM).

\bibliographystyle{unsrtnat}
\bibliography{references}
\newpage
\section{Appendix}

\begin{figure}[h!]
    \centering
    \includegraphics[width=0.75\linewidth]{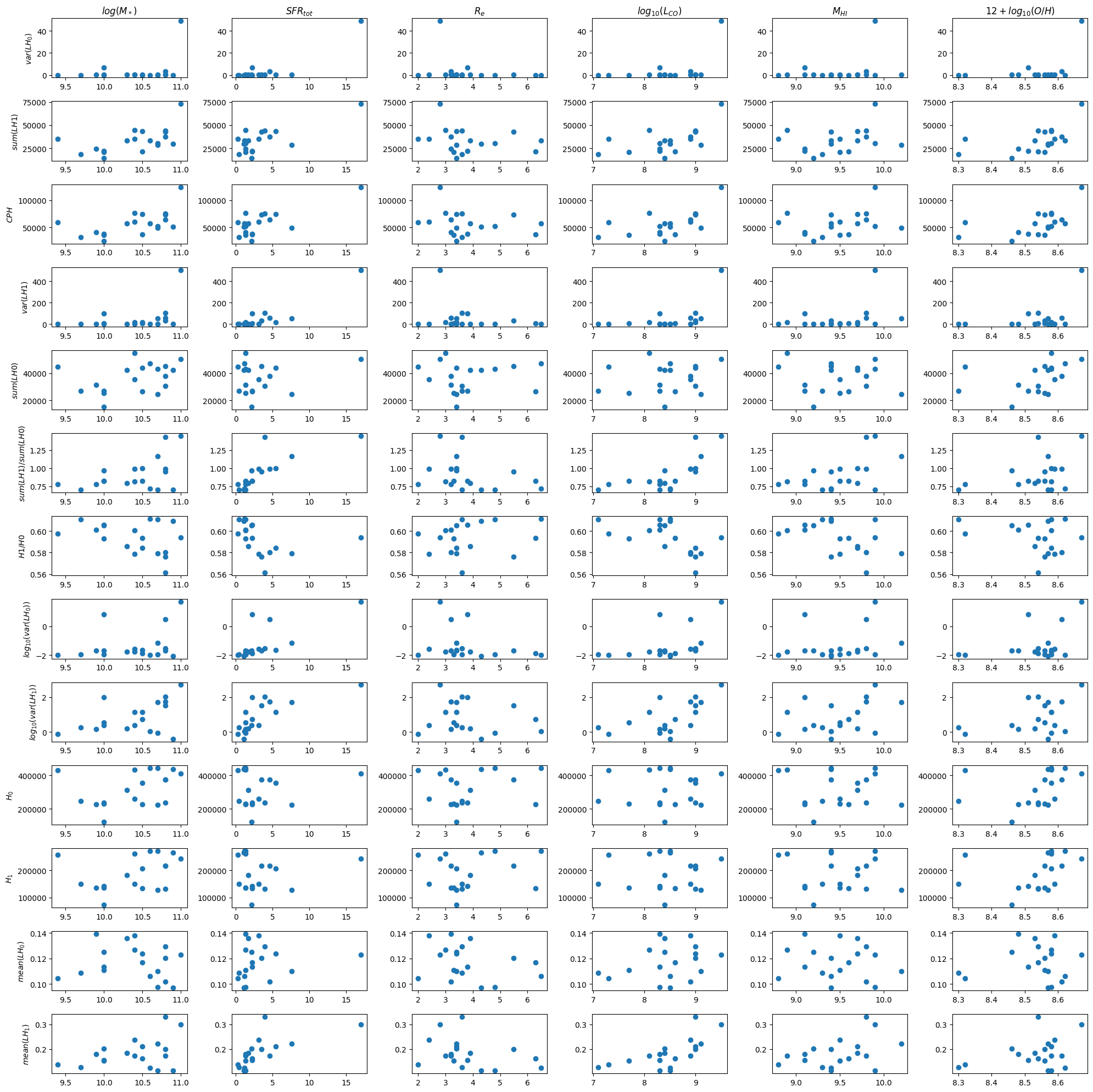}
    \caption{Table of plots of topological vs physical magnitudes.}
    \label{topology_vs_physics_img_table}
\end{figure}

\begin{figure}[h!]
    \centering
    \includegraphics[width=0.75\linewidth]{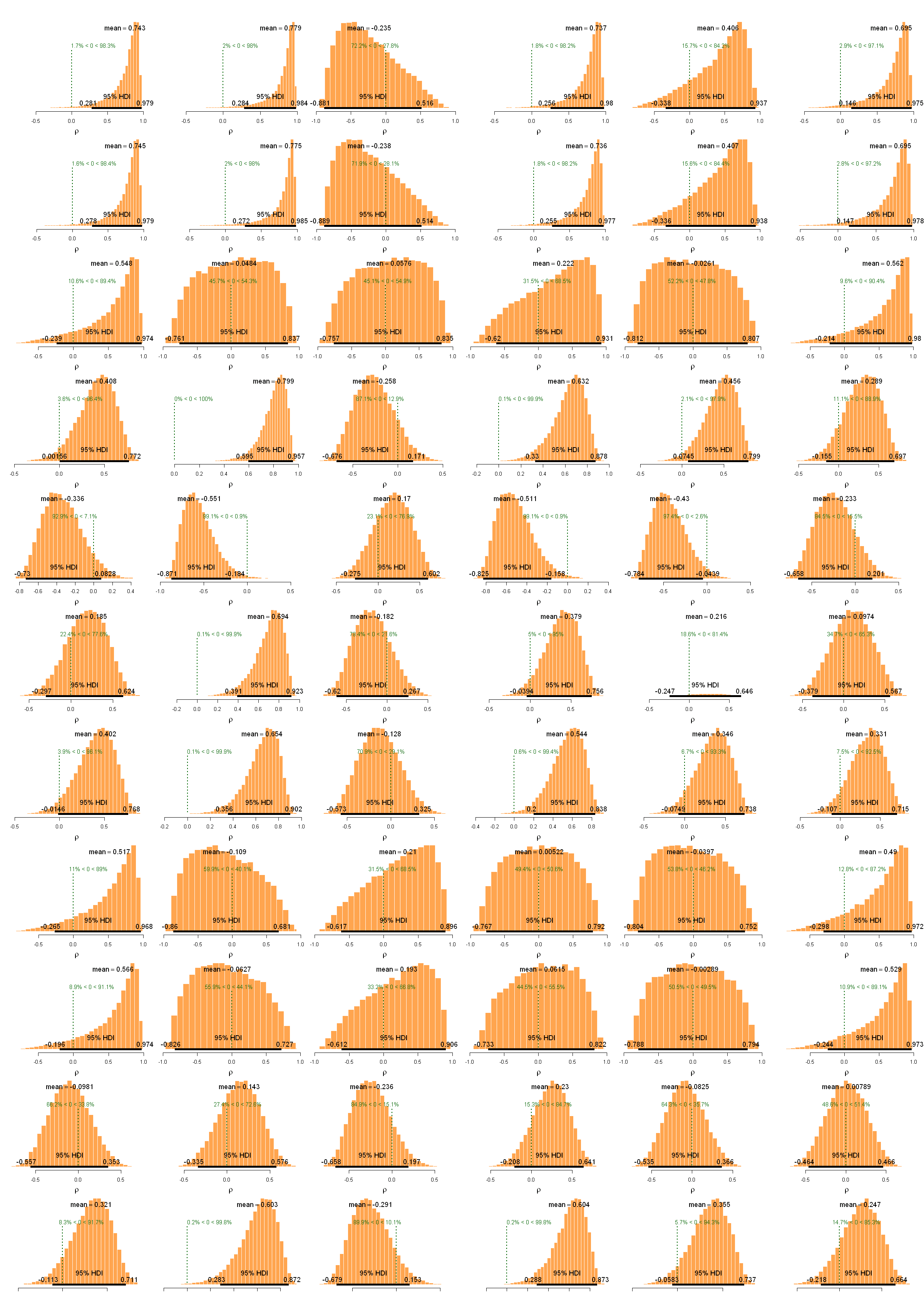}
    \caption{Regression posteriors}
    \label{regression_posteriors}
\end{figure}
\begin{figure}[h!]
    \centering
    \includegraphics[width=0.75\linewidth]{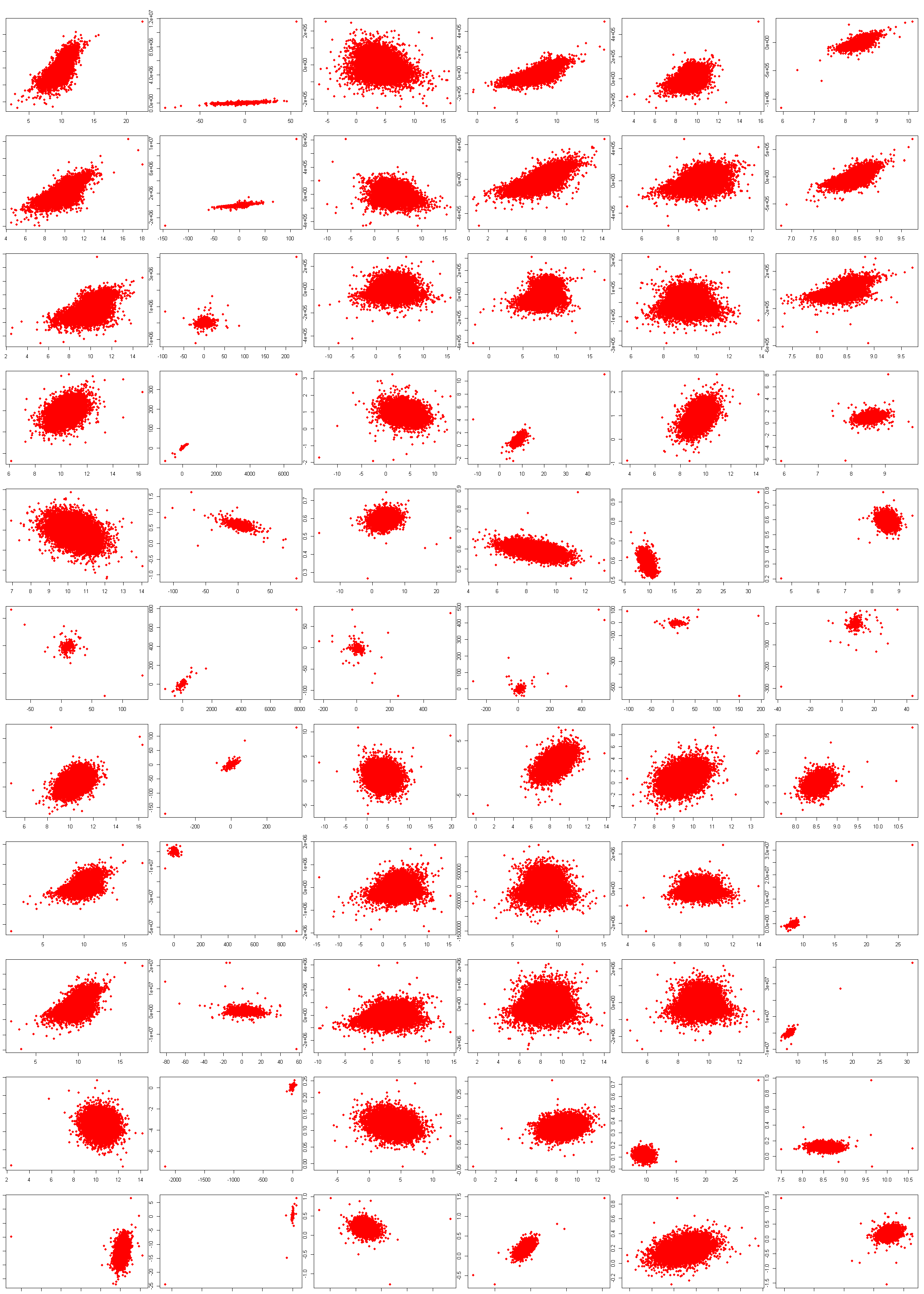}
    \caption{Posterior predictive distributions}
    \label{DPPs}
\end{figure}
\begin{figure}[h!]
    \centering
    \includegraphics[width=0.75\linewidth]{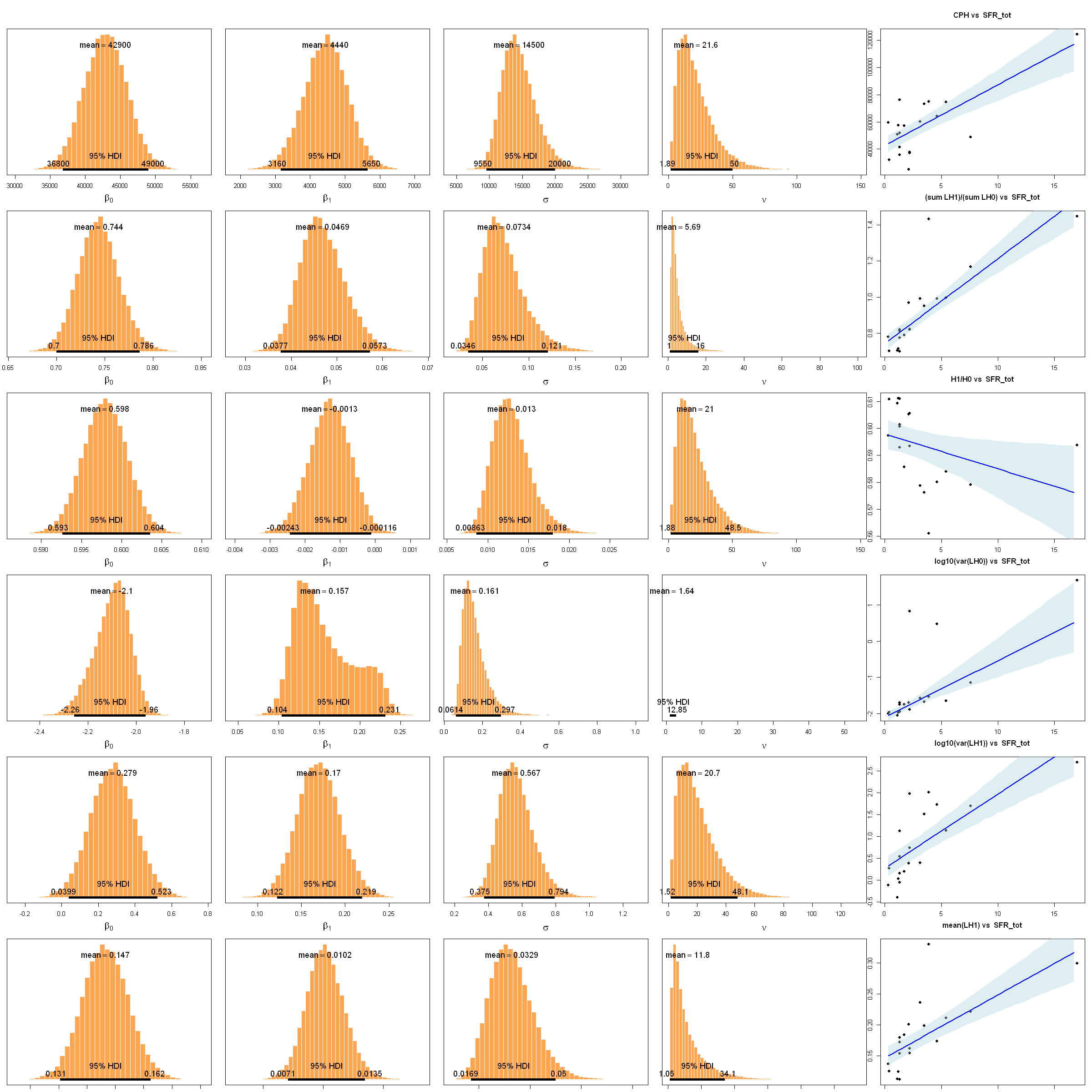}
    \caption{Linear fits. Posteriors and plot with HDI in pale blue and the mean in darker blue.}
    \label{linear_fits}
\end{figure}







\end{document}